\documentclass[twoside,12pt]{article}
\usepackage{epsfig}
\usepackage{amsmath,amsfonts}

\newcommand{\be}{\begin{equation}}
\newcommand{\ee}{\end{equation}}
\newcommand{\bea}{\begin{eqnarray}}
\newcommand{\eea}{\end{eqnarray}}

\begin{document}

\title{Symmetry energy effects on isovector properties of neutron rich nuclei with a Thomas-Fermi approach  }

\author{M.C. Papazoglou and Ch.C. Moustakidis \\
$^{}$ Department of Theoretical Physics, Aristotle University of
Thessaloniki, \\ 54124 Thessaloniki, Greece }

\maketitle

\begin{abstract}
We employ a variational method, in the framework of the Thomas-Fermi approximation, to study the effect of the symmetry
energy on the neutron skin thickness and the symmetry energy
coefficients of various neutron rich nuclei. We concentrate our
interest on $^{208}$Pb, $^{124}$Sn, $^{90}$Zr, and  $^{48}$Ca,
although the method can be applied in the totality of medium and
heavy neutron rich nuclei. Our approach has the advantage that the
isospin asymmetry function $\alpha(r)$, which is the key quantity
to calculate isovector properties of various nuclei, is directly
related with the symmetry energy as a consequence of the
variational principle. Moreover,  the Coulomb interaction is
included in a self-consistent way and its effects can be separated
easily from the nucleon-nucleon interaction. We confirm, both
qualitatively and  quantitatively,  the strong dependence of the
symmetry energy on the various isovector properties for the
relevant nuclei, using  possible constraints between the slope and
the value of the symmetry energy at the saturation density.
\\
\\
PACS number(s): 21.65.Ef, 21.65.Mn, 21.65.Cd, 21.10.Gv
\end{abstract}

\section{Introduction}
The nuclear symmetry energy (SE) is the basic regulator of the
isospin properties of the neutron rich nuclei
\cite{Danielewicz-02,Lattimer-07,Li-08,Steiner-05,Lattimer-012,Vretenar-05,Sammarruca-13,Giuliani-014}.
It is expected to affect the neutron skin thickness, the
coefficient of the asymmetry energy in Bethe-Weizsacker formula,
e.t.c. In addition, the density dependence of the SE is the  main
ingredient of the equation of state of neutron rich nuclear
matter. Actually there is a variety of neutron star properties
which are sensitive to SE, that is the maximum mass value and the
corresponding radius, the onset of the direct Urca process, the
crust-core transition density and pressure e.t.c.
\cite{Lattimer-07,Haensel-07}

Recently, there is an extended theoretical
\cite{Horowitz-14,Danielewicz-03,Danielewicz-09,Danielewicz-13,Lattimer-013,Moller-012,Ono-03,Brown-2000,
Brown-2001,Centelles-07,Centelles-09,Centelles-10,Warda-09,Vinas-03,
LWChen-05,Furnstahl-02,Kanzawa-09,Oyamatsu-07,Sammarruca-09,Agrawal-010,Agrawal-012,
LWChen-011,Mei-012,JLiu-013,Fattoyev-012,Fattoyev-013,Zhang-013,Kortelainen-013,Singh-013,Paar-013,Mekjiian-85,
Bodmer-03,Denisov-02,Prassa-010,Wolter-09,Gaidarov-011,Gaidarov-012,
Reinhard-010,Sharma-09,Blocki-013,Erler-013,Moustakidis-07,
Moustakidis-012,Psonis-07,Fan-014,Inakura-013,Cozma-013,Fattoyev-013,Mallik-013,Xu-013,Steiner-012,Souza-09,Drischler-014}
and experimental
\cite{Tsang-09,Tsang-012,Klimkiewicz-07,Abrahamyan-012,Horowitz-012,
Tarbert-013,Trzcinska-01,Shetty-04,Shetty-07,Marini-013,Veselsky-013}
interest to constrain the slope of the symmetry energy $L$ close
to the value of the saturation density $\rho_0$ of nuclear matter.
Both, theoretical and experimental efforts are focused on the
study of a possible correlation of $L $ with various nuclear
properties including the neutron skin thickness, the dipole
polarizability and the pygmy dipole resonance of various neutron
rich nuclei as well as  the analysis of heavy ion collision data.
Additionally, isobaric analog states,  nuclei mass formula data
and also neutron star observation data are also elaborated.

However, the experimental data for the SE still remain limited and
only for low values of density ($\rho < \rho_0$) are accurately
constrained. From the theoretical point of view there is an effort
to constrain the trend of SE, even for low values of density, from
finite nuclei properties and to extrapolate in a way to densities
related to neutron stars equation of state (up to $\simeq 5
\rho_0$). In any case, the constraints of $L$ or in general the
density dependence of  SE, even for low values of $\rho$, are very
important for astrophysical applications. For example the
transition density and pressure between the crust and the core in
a neutron star are expected to lie close to the half values of the
saturation density $\rho_0$ and consequently similar to the finite
nuclei interior densities
\cite{Lattimer-07,Li-08,Steiner-05,Lattimer-012}.

The structure of a  heavy nucleus is a result of the interplay
between the strong short range nuclear forces and long range
Coulomb interaction. However, in order to exhibit the isovector
character of nuclear forces, we have to focus mainly  on heavy and
additional neutron rich nuclei. Furthermore it is well known that
the energy density formalism  is able to reproduce properties of
finite nuclei including mainly the bulk properties, namely the
binding energy as well as the size and shape of the mass and
charge distributions
\cite{Erler-013,Bethe-68,Ring-80,Myers-74,Baldo-013,Niksic-011}.
The Thomas-Fermi model, which has been applied previously with success  for the study of main properties of heavy nuclei, is the main framework of the present study. More precisely, we employ a variational approach,  based on the Thomas-Fermi approximation, by suitably constructing an  energy density functional, and solving the derived Euler-Lagrange equation.
Special attention is devoted to the contribution of
the nuclear symmetry energy and  the self-consistent treatment of
the Coulomb interaction. The symmetry energy is suitably
parameterized. Actually the present approach can be easily
extended to include more complicated expressions for the symmetry
energy as well as for the energy of the symmetric nuclear matter.

The key quantity of the present study is the isospin asymmetry
function $\alpha(r)=(\rho_n(r)-\rho_p(r))/\rho(r)$ (where
$\rho_n$, $\rho_p$ and $\rho=\rho_n+\rho_p$ are the neutron,
proton  and total number densities respectively). The method has
the advantage that the asymmetry function $\alpha(r)$ is directly
related with the symmetry energy as a consequence of the
variational principle. It is expected that the various isovector
properties of nuclei (neutron skin thickness, symmetry energy
coefficient e.t.c.) depend on the trend of the symmetry energy for
densities close to the interior of the nucleus. The motivation of
the present work is twofold. Firstly we tried to construct a
self-consistent and easily applicable density functional method to
study the effect of the symmetry energy on the isovector structure
properties  of medium and heavy neutron rich nuclei. Secondly, our
aim is, if it is possible, to combine our theoretical estimation
with the relevant experimental or empirical data in order to
suggest constraints on the density dependence of the symmetry
energy for densities close to those of the interior of  finite
nuclei.

The article is  organized as follows. In Sec.~II we review the
density functional method  and the variational approach employed
for calculating the bulk properties of various neutron rich
nuclei. The results are presented and discussed in Sec.~III, while
Sec.~IV summarizes the present study.

\section{Energy density functional and variational approach }
According to the empirical Bethe-Weizsacker formula the binding
energy of a finite nucleus with A nucleons and atomic number Z is
given by
\begin{equation}
B(A,Z)=-a_VA+a_SA^{2/3}+a_C
\frac{Z(Z-1)}{A^{1/3}}+a_A\frac{(N-Z)^2}{A}+\Delta E_{mic}.
\label{B-W}
\end{equation}
The  first  term corresponds to the volume effect, the second is
the surface term, the third one  takes into account the Coulomb
repulsion of the protons, while the fourth is the symmetry energy
term. Finally, the last term corresponds to other factors
including the pairing interaction e.t.c. Using fits of known
masses to this equation one can determine the corresponding
coefficients $a_V$, $a_S$, $a_C$ and $a_A$.

The energy density functional is a natural extension of the above
formula, where now the total energy is a functional of the proton
and neutron densities and consists of  terms corresponding  with
those appearing in relation (\ref{B-W}). The minimization of the
total energy defines the related  densities and consequently the
contribution of each term separately. In the present work we apply
the energy density formalism, where the total energy of  finite
nuclei is a functional of the total density $\rho(r)$ and the
isospin asymmetry function $\alpha(r)$, that is
\begin{equation}
E[\rho(r),\alpha(r)]=\int_{\cal V}{\cal
E}\left(\rho(r),\alpha(r)\right)d^3r, \label{EDF-1}
\end{equation}
where ${\cal E}(r)$ is the local energy density. The integration
is performed over the total volume ${\cal V}$ occupied by the
relevant nuclei.

Now we consider the functional
\begin{equation}
E[\rho,\alpha]=\int_{{\cal V}} \left[
\epsilon_{ANM}(\rho(r),\alpha(r))+F_0|\nabla\rho(
r)|^2+\frac{1}{4}\rho(1-\alpha)V_c(r) \right] d^3r. \label{funt-1}
\end{equation}
The first ingredient of the functional,
$\epsilon_{ANM}(\rho(r),\alpha(r))$, corresponds to the energy
density of the asymmetric nuclear matter given by the expression
\begin{equation}
\epsilon_{ANM}(\rho,\alpha)=\epsilon_{SNM}(\rho)+\alpha^2\rho
S(\rho), \label{En-1}
\end{equation}
where  $\epsilon_{SNM}(\rho,\alpha)$ is the energy density of
symmetric nuclear matter and $S(\rho)$ is the symmetry energy per
particle of nuclear matter.

The second term $F_0|\nabla\rho( r)|^2$ is the gradient term
corresponding to the contribution originating from the finite size
character of the density distribution with $F_0$ being a parameter
in  the interval $(66-72)$ MeV. In the present work we consider
that $F_0=70$ MeV.

The third term corresponds to the Coulomb energy density where the
Coulomb potential $V_c(r)$ is defined as
\begin{equation}
V_c(r)=\frac{e^2}{2}\int\frac{\rho(r')(1-\alpha(r'))}{|{\bf
r}-{\bf r}'|}d^3r', \label{Coul-pot-1}
\end{equation}
and must satisfy also the Poisson equation
\begin{equation}
\nabla^2V_c(r)=4\pi
e^2\left(\frac{1}{2}(1-\alpha(r))\right)\rho(r). \label{Pios-1}
\end{equation}
Eq.~(\ref{Pios-1}) is used to check the convergence of the
iteration process involved in such a kind of calculations. Finally
the density $\rho(r)$ and the asymmetry function $\alpha(r)$ must
obey the following constraints
\begin{equation}
\int\rho(r) d^3r=A, \qquad \int \alpha(r)\rho( r)d^3r=N-Z.
\label{constr-1}
\end{equation}
The functional (\ref{funt-1}) and the constraints (\ref{constr-1})
after some algebra are written as
\begin{equation}
E[\rho,\alpha]=4\pi \int_0^{\infty} r^2 \left[
\epsilon_{ANM}(\rho(r),\alpha(r))+F_0\left(\frac{d\rho}{dr}\right)^2+\frac{1}{4}\rho(1-\alpha)V_c(r)
\right] dr \label{funt-3-coul}
\end{equation}
and
\begin{equation}
4\pi \int_0^{\infty} r^2 \rho(r) dr=A, \qquad 4\pi \int_0^{\infty}
r^2 \alpha(r)\rho(r)dr=N-Z. \label{constr-2-coul}
\end{equation}
Eqs.~(\ref{funt-3-coul}) and (\ref{constr-2-coul})  constitute a
variational  problem with constraints while the  Lagrangian
density is given by
\begin{equation}
{\cal L}=4\pi r^2 \left(
\epsilon_{ANM}(\rho,\alpha)+F_0\left(\frac{d\rho}{dr}\right)^2+\frac{1}{4}\rho(1-\alpha)V_c(r)
\right)-\lambda_1 4\pi r^2 \rho-\lambda_2 4\pi r^2 \alpha \rho,
\label{Lagr-1-c}
\end{equation}
In Eq.~(\ref{Lagr-1-c}) $\lambda_1$ and $\lambda_2$ are the
Lagrange multipliers. The two corresponding Euler-Lagrange
equations for $\rho(r)$ and $\alpha(r)$  are defined as follows:
\begin{equation}
\frac{\partial {\cal L}}{\partial
\rho}-\frac{d}{dr}\left(\frac{\partial {\cal L}}{\partial \rho'}
\right)=0, \label{E-L-1-c}
\end{equation}
\begin{equation}
\frac{\partial {\cal L}}{\partial
\alpha}-\frac{d}{dr}\left(\frac{\partial {\cal L}}{\partial
\alpha'} \right)=0. \label{E-L-1-c}
\end{equation}
We find easily that
\begin{equation}
\frac{\partial {\cal L}}{\partial \rho}=4\pi r^2
\left[\frac{\partial \epsilon_{SNM}(\rho)}{\partial \rho}+\alpha^2
\left(S(\rho)+\rho\frac{\partial S(\rho)}{\partial \rho}
\right)+\frac{1}{4}(1-\alpha)V_c(r) -\lambda_1 -\lambda_2 \alpha
\right], \label{EL-3-c}
\end{equation}
\begin{equation}
\frac{\partial {\cal L}}{\partial \rho'}=8\pi r^2F_0\rho',
\label{EL-3-c}
\end{equation}
\begin{equation}
\frac{d}{dr}\left(\frac{\partial {\cal L}}{\partial \rho'}
\right)=8\pi F_0r^2\rho''+16\pi F_0 \rho'r. \label{EL-4-c}
\end{equation}
Also we have
\begin{equation}
\frac{\partial {\cal L}}{\partial \alpha}=4\pi r^2\left[2\alpha
\rho S(\rho)-\frac{1}{4}\rho V_c(r)-\lambda_2  \rho \right]
 \label{EL-5-c}
\end{equation}
\begin{equation}
\left(\frac{\partial {\cal L}}{\partial \alpha'} \right)=0.
\label{EL-6-c}
\end{equation}
The first Euler-Lagrange equation gives
\begin{equation}
\rho''+\frac{2\rho'}{r}-\frac{1}{2F_0}\left[\frac{\partial
\epsilon_{SNM}(\rho)}{\partial \rho}+\alpha^2
\left(S(\rho)+\rho\frac{\partial S(\rho)}{\partial \rho}
\right)+\frac{1}{4}(1-\alpha)V_c(r) -\lambda_1 -\lambda_2 \alpha
\right]=0 \label{dif-1-c}
\end{equation}
and the second one
\begin{equation}
\alpha(r) =\frac{V_c(r)}{8S(\rho)}+\frac{\lambda_2}{2S(\rho)}
=\frac{1}{8S(\rho)}\left(\frac{}{} V_c(r)+4\lambda_2
\right).\label{dif-2-c}\end{equation}
The asymmetry function $\alpha(r)$ obeys the constraints $0\leq
\alpha(r) \leq 1$. However, the expression (\ref{dif-2-c}) does
not ensure the above constraints, since for high values of $r$
(low values of $\rho(r)$ and consequently $S(\rho)$) $\alpha(r)$
increases very fast and there is a cut-off radius, $r_c$ where
$\alpha(r_c)=1$ and also $\alpha(r \geq r_c)\geq 1$.

In order to overcome this unphysical behavior of $\alpha(r)$ we
use the assumption
\begin{eqnarray}
\alpha(r)&=&\left\{
\begin{array}{ll}
\frac{1}{8S(\rho)}\left(\frac{}{} V_c(r)+4\lambda_2
\right), \qquad r \leq r_c       &          \\
\\
1, \qquad r\geq r_c.  &  \
                              \end{array}
                       \right.
\label{resip-2}
\end{eqnarray}
Accordingly the proton and neutron density distributions take the
form
\begin{eqnarray}
\rho_p(r)&=&\left\{
\begin{array}{ll}
\frac{1}{2}\rho(r)\left(1-\alpha(r)\right), \qquad r \leq r_c       &          \\
\\
0, \qquad r \geq r_c.  &  \
                              \end{array}
                       \right.
\label{resip-3}
\end{eqnarray}
\begin{eqnarray}
\rho_n(r)&=&\left\{
\begin{array}{ll}
\frac{1}{2}\rho(r)\left(1+\alpha(r)\right), \qquad r \leq r_c       &          \\
\\
\rho(r), \qquad r \geq r_c.  &  \
                              \end{array}
                       \right.
\label{resip-4}
\end{eqnarray}
The Lagrange multiplier $\lambda_2$ is found from the
normalization condition
\begin{equation}
\int_{{\cal V}} \alpha(r)\rho(r)d^3r=N-Z, \label{resip-5}
\end{equation}
where the integration is performed over the total volume occupied
by the  specific nucleus considering that $\alpha(r)$ is given by
(\ref{resip-2}). After a straightforward algebra we get
\begin{equation}
\lambda_2=2\left(\int_{{\cal V}_c} \rho(r)d^3r
-\frac{e^2}{8}\int_{{\cal V}_c}\frac{V_c(r)\rho(r)}{S(\rho)}d^3r
-2Z\right)\left(\int_{{\cal
V}_c}\frac{\rho(r)}{S(\rho)}d^3r\right)^{-1}, \label{LM-2}
\end{equation}
where ${\cal V}_c$ is the part of the  spherical volume of the
nucleus  for  the radius $r_c$.  The cut-off radius $r_c$, which
reflects the combined effect of the symmetry energy and Coulomb
energy on the asymmetry function $\alpha(r)$ according to
expression (\ref{dif-2-c}), easily can be determined  by solving
the equation
\begin{equation}
\alpha(r_c)=1. \label{rc-1}
\end{equation}
Finally from the equations (\ref{resip-2}), (\ref{LM-2}) and
(\ref{rc-1}) we see that the  asymmetry function $\alpha(r)$ is
\begin{equation}
\alpha(r)=\frac{1}{S(\rho)}\left( \frac{V_c(r)-V_c(r_c)}{8}
+S(\rho_c)\right), \qquad r \leq  r_c, \label{asym-final}
\end{equation}
and $\alpha(r)=1$ for $r \geq r_c$. In Eq.~(\ref{asym-final}) one
can see clearly exhibited the interplay between the long-range
Coulomb interaction and the short-range isovector part of the
nuclear forces. Actually $S(\rho)$ affects $\alpha(r)$ in a
twofold manner: a) directly via the term $S(\rho_c)/S(\rho)$ and
b) indirectly since the $V_c(r)$ according to
Eq.~(\ref{Coul-pot-1}) is a functional of $\alpha(r)$. In the
simplified case where $V_c(r)$ is excluded, the asymmetry function
is given by the simple formula
\begin{equation}
\alpha(r)=\frac{S(\rho_c)}{S(\rho)}. \label{ar-simp}
\end{equation}
The Coulomb potential, given by Eq.~(\ref{Coul-pot-1}) due to the
discontinuity behavior of the proton density distribution
(Eq.~(\ref{resip-3})) is decomposed in two parts as follow
\begin{equation}
V_c^A(r)=2\pi e^2  \left[ \frac{1}{r}\int_{0}^r(1-\alpha(r'))
\rho(r')r'^2dr'+\int_{r}^{r_c}\left(1-\alpha(r')\right)\rho(r')
r'dr' \right], \qquad r \leq r_c \label{Vc-A-1}
\end{equation}
\begin{equation}
V_c^B(r)=\frac{2\pi e^2}{r} \int_{0}^{r_c}(1-\alpha(r'))
\rho(r')r'^2dr', \qquad r \geq r_c. \label{Vc-B-1}
\end{equation}
Actually one has to solve self-consistently Eqs.~(\ref{dif-1-c})
and~(\ref{dif-2-c}) with the corresponding constraints
~(\ref{constr-2-coul}). In the present work, in order to avoid the
complication due to the differential equation ~(\ref{dif-1-c}) we
employ a variational method where use is made of an appropriate
trial function for $\rho(r)$. This method, as pointed out by
Brueckner {\it et. al.}
\cite{Brueckner-68,Brueckner-69,Brueckner-71,Buchler-71,Lombard-73},
provides a convenient tool in seeking approximate solution for
heavy nuclei. There is a variety of trial density distribution
functions suitably parameterized to describe light, medium and
heavy nuclei. In the present study we consider the trial function
given by the Fermi type formula
\begin{equation}
\rho(r)=\frac{n_0}{1+\exp[(r-d)/w]}. \label{Fermi-1-1}
\end{equation}
In addition, for the basic ingredients of the energy functional
(\ref{funt-3-coul}) we consider a model where the energy of the
symmetric nuclear matter is given by \cite{Myers-98}
\begin{equation}
\epsilon_{SNM}(\rho)=\rho T_o\left(au^{2/3}-bu+cu^{5/3}   \right),
\qquad u=\rho/\rho_0, \label{MN-model-Mayers}
\end{equation}
where $T_o=37.0206$ MeV and $\rho_0=0.16144$ fm$^{-3}$ (the
saturation density). The corresponding constants are:
$a=-0.08203$, $b=0.97342$ and $c=0.61687$.

The symmetry energy $S(\rho)$ can be suitably expanded around the
saturation density $\rho_0$ as follows
\begin{equation}
S(\rho)=S(\rho_0)+L\delta+\frac{K_{sym}}{2!}\delta^2+{\cal
O}(\delta^3), \label{Esym-taylor-1}
\end{equation}
where $S(\rho_0)$ is  the value of the symmetry energy at the
saturation density and $\delta=\frac{\rho-\rho_0}{3\rho_0}$. The
coefficient $L=3\rho_0\frac{dS(\rho)}{d\rho}|_{\rho=\rho_0}$ is
related with the slope of the symmetry energy at $\rho_0$, while
the coefficient $K_{sym}$ is given by
$K_{sym}=9\rho_0^2\frac{d^2S(\rho)}{d\rho^2}|_{\rho=\rho_0}$.

There are various suggested expressions for the symmetry energy in
the literature. Here we employ the simple parameterization
\begin{equation}
S(\rho)=S(\rho_0)\left(\frac{\rho}{\rho_0}
\right)^{\gamma}=Ju^{\gamma}, \quad S(\rho_0)=J. \label{esym-1}
\end{equation}
Obviously, in this case the parameter $\gamma$ is related  with
both the slope $L$ and $J$ by the expression
\begin{equation}
\gamma=\frac{L}{3J}. \label{g-1}
\end{equation}
It is worth pointing out that for $\gamma <1 $ a smaller value of
$\gamma $ gives a stiffer $S(\rho)$ for $\rho<\rho_0$ while, for
$\rho> \rho_0$ the  higher the value of $\gamma$ the  stiffer is
$S(\rho)$. Finally, the symmetry energy density $s(\rho)$ is given
by
\begin{equation}
s(\rho)=\rho J u^{\gamma}. \label{esym-2}
\end{equation}
Now the total energy density of the asymmetric nuclear matter is
\begin{equation}
\epsilon_{ANM}(\rho,\alpha)=\rho T_o\left(au^{2/3}-bu+cu^{5/3}
\right)+\alpha^2 \rho J u^{\gamma}.\label{appl-1}
\end{equation}

For each specific set of the Fermi type distribution parameters
$n_0$, $d$, and $w$ and a given symmetry energy $S(\rho)$, we
calculate the asymmetry density $\alpha(r)$ and the total energy
of the specific nucleus. The set of the density distribution
parameters is adjusted in order to find the corresponding minimum
value of the total energy given now by the integrals
\begin{eqnarray}
E[\rho(r);\gamma]&=&4\pi \int_0^{r_c} r^2 \left(
\epsilon_{ANM}(\rho(r),\alpha(r))+F_0\left(\frac{d\rho}{dr}\right)^2+\frac{1}{4}\rho(r)(1-\alpha(r))V_c(r)
\right) dr  \\
&+&4\pi \int_{r_c}^{\infty} r^2 \left(
\epsilon_{ANM}(\rho(r),1)+F_0\left(\frac{d\rho}{dr}\right)^2
\right) dr. \nonumber
 \label{resip-22}
\end{eqnarray}
After finding the  density $\rho(r)$ and asymmetry function
$\alpha(r)$ which minimizes the total energy, all the relevant
quantities are easily calculated.

One possibility is  to calculate the symmetry energy coefficient
$a_{A}$, defined in Bethe-Weizsacker formula  via the local
density approximation. In this approach $a_A$ is defined by the
integral
\begin{equation}
a_A=\frac{A}{(N-Z)^2}\int\rho(r) S(\rho)\alpha^2(r) d^3r.
\label{sym-coef-1}
\end{equation}
Definition (\ref{sym-coef-1}) shows explicitly the direct strong
dependence of $a_A$ on  the symmetry energy $S(\rho)$ and the
asymmetry function $\alpha(r)$. Actually, according to the present
study,  the total integral is split in two parts as follows
\begin{equation}
a_A=\frac{A}{(N-Z)^2}\left(\int_{{\cal V}_c}\rho(r)
S(\rho)\alpha^2(r) d^3r+\int_{{\cal V}>{\cal V}_c}\rho(r) S(\rho)
d^3r\right). \label{resip-24}
\end{equation}
One of the most important quantities concerning the isovector
character of the nuclear forces is the neutron skin  thickness
defined as
\begin{equation}
R_{skin}=R_n-R_p,\label{resip-25}
\end{equation}
with $R_n$ and $R_p$ the neutron and proton and  radii
respectively defined as
\begin{equation}
R_n=\left(\frac{1}{N}\int r^2 \rho_n(r) d^3r\right)^{1/2}, \qquad
R_p=\left(\frac{1}{Z}\int r^2 \rho_p(r) d^3r\right)^{1/2}.
\label{rn-rp-1}
\end{equation}
In the framework of the present approach they are given
respectively by the expressions
\begin{equation}
R_n=\left[\frac{1}{N}\left(\int_{{\cal V}_c} r^2
\frac{1}{2}\rho(r)\left(1+\alpha(r)\right) d^3r+\int_{{\cal V} >
{\cal V}_c} r^2 \rho(r) d^3r \right)\right]^{1/2} \label{resip-26}
\end{equation}
and
\begin{equation}
R_p=\left(\frac{1}{Z}\int_{{\cal V}_c} r^2
\frac{1}{2}\rho(r)\left(1-\alpha(r)\right) d^3r\right)^{1/2}.
\label{resip-27}
\end{equation}
Actually, $R_{skin}$ is not directly dependent on $S(\rho)$,
compared to the case of $a_A$, but indirectly via the dependence
of $\alpha(r)$. However,  recent studies conjecture that
$R_{skin}$ is a strong indicator of the isospin character of the
nuclear interaction expected to be strongly correlated with the
symmetry energy slope $L$ and the value $J$ or in general with the
values of the symmetry energy close to the saturation density.

\section{Results and Discussion}
We employ a variational approach  to study the effect of the
symmetry energy on isovector properties of various medium and
heavy nuclei. The method, even its simplicity, has the advantage
that the dependence of the asymmetry function $\alpha(r)$ on the
symmetry energy, and  the Coulomb potential, are introduced
explicitly. More specifically, the total energy density of the
nucleus consists of the nuclear  and Coulomb contributions. The
nuclear term consists of two parts i.e. the symmetric nuclear
matter and also  the asymmetry energy one. In order to be able to
study the effects of the symmetry energy we parameterized suitably
the related expression. It is noted that the results obtained in the present study
are based on the assumption given in Eq.~(\ref{esym-1}) where accordingly the main parameters
$L$ and $J$ are linearly related, that is  $L=3\gamma J$.   We study the dependence of the neutron
skin thickness $R_{skin}$ and the asymmetry coefficient $a_A$ on
the slope $L$ of the SE at the saturation density $\rho_0$. In the
present method the asymmetry function $\alpha(r)$ is treated as a
variational function and the total density $\rho(r)$ as a trial
function. For each $\rho(r)$ the corresponding $\alpha(r)$ and the
total energy are found. The process continues up to find the
function $\rho$ which minimizes the total energy. All the relevant
quantities, which are functionals of $\rho(r)$, $\alpha(r)$ and
$S(\rho)$, are easily calculated.

The outline of our approach is the  following: We start from the
general relation $R=r_0A^{1/3}$ which gives an averaged estimate
of the nuclear radius and accordingly we consider a Fermi form for
the total density distribution $\rho(r)$ \cite{Hasse-88}.
Afterwards, for a fixed $\rho(r)$ the asymmetry function is
rearranged accordingly so that the total energy of the nucleus is
the lowest one.

It is worth to point out, following the discussion by Brueckner
et.al. \cite{Brueckner-69} that the energy density functional
(\ref{funt-1}) breaks down at the edge of the nucleus for two
reasons. Firstly, the Thomas-Fermi approximation, which is the
basis of the present work, fails for low densities. Secondly at
the edge of the nucleus the asymmetry function $\alpha(r)$ tends
to unity and the potential contribution to the total energy
functional is not accurate. For a recent discussion on the connection of the density functional formalism with the nuclear matter equation of state and the distinct features of finite-size effect of nuclei see Ref.~\cite{Baldo-010}.   In the present work we employ a variational treatment of an  energy
functional, without any additional constraints requiring  just the
minimization of the binding energy.
That is we do not impose any additional constraints on the
functional, for example to reproduce accurately  the proton radii
e.t.c. This  approach will be suitable
if we intend to  impose stronger constraints on the values of $L$
and $J$ and might be of interest for future work.
However,  the main  motivation  of the present  work   is not to
find the suitable energy density functional to reproduce simultaneously
the experimental values of proton radii and energy but to focus on
the symmetry energy effects on neutron rich nuclei properties.

In Fig.~1, the symmetry energy versus the total density is
plotted, according to Eq.~(\ref{esym-1}) for various values of the
slope parameter $L$. It is noted that  lower values of $L$, for low values of densities ($\rho < \rho_0$),
correspond to higher values of $S(\rho)$. This behavior of
$S(\rho)$ is well reflected  on the values of the total binding
energy $E_{tot}$ and the asymmetry function $\alpha(r)$. More
precisely, higher values of $L$ lead to lower contribution of the
$S(\rho)$ on the total binding energy and consequently the
nucleons become more bound. For example in Table I are presented
the results (concerning the total binding energy $E_{tot}$, the
proton $R_p$ and neutron $R_n$ rms radii, the neutron skin
$R_{skin}$ and the asymmetry coefficient $a_A$) for the nucleus
$^{208}$Pb for the case $J=30$ MeV and for $10\ {\rm MeV} \leq L
\leq 100\ {\rm MeV} $. In addition, since the values of the
density distribution inside the nucleus are lower than the value
of the saturation density $\rho_0=0.16$ fm$^{-3}$, it is concluded
that the isovector properties of  nuclei are related with the
trend of the symmetry energy in the region $0<\rho < \rho_0$ and
vice-versa, that is the experimental isovector measurements  give
information for the lower part of the SE.

In Fig.~2 we plot the density distributions (total, proton and
neutron) as well as the corresponding asymmetry function
$\alpha(r)$ for various values of $L$ and  two nuclei ($^{208}$Pb
and $^{48}$Ca.) The softness symmetry energy (higher values of
$L$) shift the neutron distribution to the outer part of the
nucleus, while at the same time it concentrates deeper the
protons. This is clearly reflected both on the corresponding
values of $R_p$ and $R_n$ as well as on the neutron skin
$R_{skin}$ (see also table I). The effects of the symmetry energy
is even more pronounced on the trend of the asymmetry function.
Higher values of $L$ shift the cut-off radius $r_c$ to even
lower values increasing dramatically the neutron skin and forming
a kind of {\it neutron halo} inside the nucleus. It is obvious
from the above analysis that $S(\rho)$ and consequently, according
to expression (\ref{asym-final}), the asymmetry function
$\alpha(r)$ acts as a {\it regulator} on the proton and neutron
distributions in order to minimize, in every case, the total
energy of the nucleus. On the other hand, and accordingly (see also
expression (\ref{asym-final})) the Coulomb potential $V_c(r)$ acts
inversely, compared to $S(\rho)$ and its main effect is to shift
the proton distribution to the outer part of the nucleus.
Actually, the interplay between the long range coulomb forces, the
nuclear forces and mainly the  isovector part of nuclear forces is
responsible for the creation of the neutron skin thickness.
However, although the Coulomb contribution is well defined, the
contribution of the symmetry energy still remains an open problem
even for low values of densities.

Fig.~3(a) displays  the neutron skin $R_{skin}$ as a function of $L$
for various values of $J$ for $^{208}$Pb. The most
striking feature is, in all cases, the strong dependence of
$R_{skin}$ on $L$. For a  comparison, we include for the case of
$^{208}$Pb an approximate linear dependence
\begin{equation}
R_{skin}({\rm fm})=0.101+0.00147 \ L \ ({\rm MeV}),
\label{approx-Roca}
\end{equation}
established by Centelles {\it et. al.,} \cite{Centelles-09} using
a wide range of non-relativistic and relativistic models. It is
obvious that relation (\ref{approx-Roca}) supports a softer
dependence of $R_{skin}$  on $L$ compared to the present study.
However we note that we present a systematic study of the effects
of $L$ on $R_{skin}$  and in a large range of values of $L$
without trying to reproduce for example the experimental value of
the binding energy or the charge radius of the specific nucleus.
Even in this case, we found that the  intersection between our
results and the results compatible with (\ref{approx-Roca})
corresponds to values of binding energy very close to the
experimental  for the specific nuclei.

Very recently the Lead Radius Experiment (PREX) at the Jefferson
Laboratory has provided the first model-independent evidence for
the existence of a neutron-rich skin in $^{208}$Pb
\cite{Abrahamyan-012,Horowitz-012}. The determined neutron skin
was $R_{skin}=0.33^{+0.16}_{-0.18}$fm. However such a large error
is not  enough to constrain the various nuclear models. In
addition the large determined neutron skin (compared to previous
experimental measurements) creates a new open problem concerning
the correlation between the nuclear equation of state of nuclear
matter and the density functional theory in finite nuclei (see for
a  pertinent discussion in \cite{Fattoyev-012,Fattoyev-013}).

In Fig.~3(b) we display the  coefficients $a_A$ as a function of $L$,
for various values of $J$. It is obvious that $a_A$ is a
decreasing function of $L$. Actually, for specific pairs of values
of $N$ and $Z$, and according to Bethe-Weizsacker formula
(\ref{B-W}), a softer $S(\rho)$ (high values of $L$) leads to a
lower value of $\alpha(r)$ (a property directly connected with the
contribution of the symmetry energy to the total energy).
Obviously, $a_A$ exhibits a mass depended $A$ behavior.

In order to impose some possible constraints on the values of $L$,
we plot  in Fig.~4(a) for each of the four nuclei $^{208}$Pb,
$^{124}$Sn, $^{90}$Zr, and  $^{48}$Ca the pairs of $L$ and $J$
consistent with the corresponding empirical values of $a_A$
determined by the formula \cite{Danielewicz-13}
\begin{equation}
a_A^{-1}=(a_V)^{-1}+(a_S)^{-1}A^{-1/3}, \label{fit-Daniel}
\end{equation}
where we use for the volume and surface coefficients $a_V=35.5$
MeV and $a_S=9.9$ MeV respectively.
 By combining Eq.~(\ref{dif-2-c}), (\ref{resip-5}) and (\ref{sym-coef-1})  we get the expression
\begin{equation}
\frac{A}{a_A}={\cal I}_1\left[1+\frac{1}{64 (N-Z)^2}({\cal I}_1{\cal I}_3-{\cal I}_2^2) \right]^{-1},
\label{A-f-1}
\end{equation}
where
\begin{equation}
{\cal I}_1=\int\frac{\rho(r)}{S(\rho)} d^3r, \quad {\cal I}_2=\int\frac{V_c(r)\rho(r)}{S(\rho)} d^3r, \quad
{\cal I}_3=\int\frac{V_c^2(r)\rho(r)}{S(\rho)} d^3r.
\label{integrals}
\end{equation}
The integral ${\cal I}_1$ is decomposed as
\begin{equation}
{\cal I}_1=\frac{A}{J}+\frac{1}{J}\int \rho(r)\left(\frac{J}{S(\rho)}-1 \right) d^3r.
\label{A-f-2}
\end{equation}
The main contribution to the integral of the right-hand  side of Eq.~(\ref{A-f-2}) originates mainly from  the surface region \cite{Danielewicz-13}.
Taking into account the surface contribution denoted by the coefficient $Q_s$,
Eq.~(\ref{A-f-1}) finally is written as
\begin{equation}
\frac{A}{a_A}=\left(\frac{A}{J\left(1+\Delta_c\right)}+\frac{A^{2/3}}{Q_s\left(1+\Delta_c\right)}\right) ,
\label{A-f-3}
\end{equation}
where $Q_s$ is given by
\begin{equation}
Q_s=J A^{2/3} \left(\int \rho(r)\left(\frac{J}{S(\rho)}-1 \right) d^3r   \right)^{-1},
\label{A-f-4}
\end{equation}
while the  contribution due to the Coulomb interaction $\Delta_c$ is
\begin{equation}
\Delta_c=\frac{1}{64 (N-Z)^2}({\cal I}_1{\cal I}_3-{\cal I}_2^2).
\label{Deltac}
\end{equation}
%
Comparing  formula (\ref{fit-Daniel}) and Eq.~(\ref{A-f-3}) it is obvious that the volume coefficient $a_V$  and the surface coefficient $a_S$  are directly related  with the value of the symmetry energy at the saturation density $J$, and the coefficient $Q_s$ respectively. In the specific  case where the Coulomb interaction  is excluded,  $a_V$ and $a_S$ are identified with $J$ and $Q_s$ respectively~\cite{Danielewicz-13}.

It is seen in Fig.~4(a) that
the set $J=34$ MeV and $L=65$ MeV reproduces very well the
empirical values of $a_A$ for almost all the medium and heavy
isotopes.

Two important features of the relation between $L$ and $J$ are
useful. First, the inequality $4.13 \leq \Delta L/\Delta J \leq
5.18$ holds approximately. This means  that a change of $1$ MeV in
the value of $J$ results in a corresponding change $4.13 \leq
\Delta L \leq 5.18$ MeV. That is according to the present approach
the accuracy on the measurements of $a_A$ and $J$ will impose
strong constraints on the values of  $L$.

Second, since the four almost linear curves are arranged very
close with a similar slope, we may conjecture that  a possible
universal dependence holds between $L$ and $J$ for nuclei at least
in the mass region A=40-200. That means that the same set of $L$
and $J$, related with nuclear symmetry energy, reproduce in a very
good accuracy the symmetry energy coefficient for medium as well
as heavy nuclei. Especially for values of $J$  and $L$ in the
region $34_{-0.2}^{+0.2}$ MeV and $65_{-1}^{+1}$ MeV respectively
the accuracy is high.

In addition, in Fig.~4(b) we present  the mass dependence of the
coefficient $a_A$ of several isotopes of Ca, Zr, Sn and Pb, for
two sets of $L$ and $J$. For a comparison we include also the
formula (\ref{fit-Daniel}). The first set ($J=32$ MeV and $L=70$
MeV) reproduces on the  average the binding energies of the
corresponding isotopes while the second set ($J=34$ MeV and $L=65$
MeV) as we mentioned above, reproduces on the average  the
empirical values of $a_A$ for the isotopes $^{208}$Pb, $^{124}$Sn,
$^{90}$Zr, and  $^{48}$Ca.

In Fig.~5 we compare the allowed pairs of $L$ and $J$ constrained
from  heavy-ion collisions and nuclear structure observable
\cite{Horowitz-14} with those found in the present approach.
Actually the present results lie inside  the intersection area
suggested by the measurements of the dipole polarizability $a_D$
as well as those found by heavy-ion collisions experiments.
However, they lie outside the interval constrained by the nuclear
masses measurements, connected with the binding energy,  but only
for $J<33$ MeV. This is due to the fact that constraints between
$L$ and $J$ are based on the adjustment of the theoretical to the
empirical value of the asymmetry coefficient $a_A$ and not on the
corresponding experimental values of the binding energies. It is
remarkable that the allowed pairs of $L$ and $J$ consistent with
our approach (see the colored lines in Fig.~5) lie at the
intersection of the bands originating from other approaches
\cite{Horowitz-14}.

Fig.~6(a) exhibits the dependence of the coefficient $a_A$ on the
asymmetry parameter $I=(N-Z)/A$ for various isotopes, for the
cases $J=32,\ L=70$ and $J=34,\ L=65$. In almost all cases there
is a soft dependence of $a_A$ on $I$ but, as expected, a strong
dependence on the value of $J$. Similarly, in Fig.~6(b) we
indicate the dependence of the neutron skin on the asymmetry
parameter $I$. The most characteristic trend  is the occurrence of
strong and linear dependence of $R_{skin}$ on $I$ that is
\begin{equation}
R_{skin}=a+b I \label{R-I}
\end{equation}
where the constants $a$ and $b$ vary in the intervals $-0.02 \leq
a \leq 0.045$ and $1.31 \leq b \leq 1.45$. Of course those
intervals are strongly dependent on the specific set of values of
$L$ and $J$. For a comparison a similar relation  suggested in
\cite{Trzcinska-01} is presented
\begin{equation}
R_{skin}=(1.01\pm 0.15)I+(-0.04\pm0.03), \label{R-I-exp}
\end{equation}
although in a number of analyzed cases the statistical errors are
rather large.

In any case, additional experimental  work is necessary to
constrain the neutron skin \cite{Horowitz-14}. In particular the
upcoming Lead Radius Experiment II  (PREX-II) promises to
determine the neutron-skin thickness of $^{208}$Pb with a
$\pm0.06$ fm accuracy while the Calcium Radius Experiment (CREX)
will provide a high precision measurement  for the neutron radius
of $^{48}$Ca with accuracy $\pm0.02$ fm \cite{Horowitz-14}. The
above measurements will be a very good test for all the
theoretical models including the present variational approach.

\section{Conclusions}
In the present work we employ  a variational method,  in the
framework of the Thomas-Fermi approximation, in order to study
the symmetry energy effects on  isovector properties of various
neutron rich nuclei. The key quantity is the asymmetry function $\alpha(r)$ naturally computed by the
variational principle. Actually, $\alpha(r)$ is a functional both
of the symmetry energy as well as  the Coulomb potential, it
contains the interplay between the long-range Coulomb forces and
short-range nuclear ones and it  defines the density distribution
of neutrons and protons. All the calculated properties are studied
as a function of the slope of the symmetry energy and the value of
the symmetry energy at the nuclear saturation density. Since, the
SE even for low values of nuclear matter is uncertain, the above
parameterization is necessary. We find that the neutron skin
thickness is very sensitive to $L$  i.e. it increases rapidly with
$L$. This is expected at least in the present approximation, since
the main ingredient of the relevant calculated integrals, the
function $\alpha(r)$ approaches unity very rapidly close to the
critical value of $r_c$ (at the surface of the proton
distribution). The above characteristic behavior of $\alpha(r)$ is
well reflected on the asymmetry coefficient $a_A$ which is a
decreasing function of $L$. In the case of $^{208}$Pb we compare
our results with those originating from additional calculations
with different models. We conclude that the present approximation
supports a stronger sensitivity of the neutron skin thickness on
$L$. However, constraining the total binding energy to be close to
the experimental one we see that our results are very close to the
mentioned empirical formula.

Our findings, from the present study, show that the experimental
knowledge of the symmetry energy at the saturation density $J$
will impose, via the values of the symmetry coefficient $a_A$,
strong constraints on $L$. More specifically, we note that the set
$J=34$, $L=65$ reproduces very well the empirical values of $a_A$
corresponding to the nuclei  under consideration. In any case,
further experimental and theoretical work is necessary for a more
detailed exploration of the effects of the symmetry energy on the
properties of finite nuclei as well as on the neutron star
structure.

\section*{Acknowledgments}
This work was supported by the A.U.Th. Research Committee under
Contract No. 89286. One of the authors (ChCM) would like to thank
P. Ring, G.A. Lalazissis, N. Paar and C.P. Panos for the fruitful
discussions.

\newpage
\begin{table}[h]
\begin{center}
\caption{The slope parameter $L$ (in MeV), the binding energy $E$
(in MeV), the neutron radius $R_n$ (in fm), the proton radius
$R_p$ (in fm), the neutron skin thickness $R_{skin}$ (in fm) and
the asymmetry coefficient $a_A$ (in MeV) calculated for a fixed
value $J=30$ MeV.  $L$ and $J$ are linearly related, that is  $L=3\gamma J$.  }
 \label{t:1}
\vspace{0.5cm}
\begin{tabular}{|c|c|c|c|c|c|}
\hline
 $L$     &   $E$     & $R_n$  &  $R_p$   & $R_{skin}$    &  $a_A$              \\
\hline
 10      &  -1581.36 &  5.629 & 5.623    & 0.006         & 28.52                                        \\
 \hline
 20      & -1593.44  & 5.659  & 5.604    & 0.055         & 26.77                                 \\
 \hline
 30      & -1606.35  &  5.695 & 5.586    & 0.109         &  24.85                                    \\
 \hline
 40      & -1620.28  & 5.723  & 5.560    & 0.163         & 22.90                                \\
 \hline
 50      &  -1633.62 & 5.756  & 5.537    & 0.219         & 20.87                                     \\
 \hline
 60      & -1646.40  & 5.780  & 5.517    & 0.263         & 19.03                                  \\
 \hline
 70      & -1658.36  & 5.803  & 5.500    & 0.303         & 17.32                                       \\
 \hline
 80      & -1669.35   & 5.816  & 5.478    & 0.338         & 15.80                                \\
 \hline
 90      &  -1679.37 & 5.828  & 5.458    & 0.370         & 14.41                                     \\
 \hline
 100     & -1688.46  & 5.846  & 5.448    & 0.398         & 13.11                               \\
 \hline
\end{tabular}
\end{center}
\end{table}

\begin{figure}
\centering \vspace{3cm}
\includegraphics[height=7.1cm,width=10.1cm]{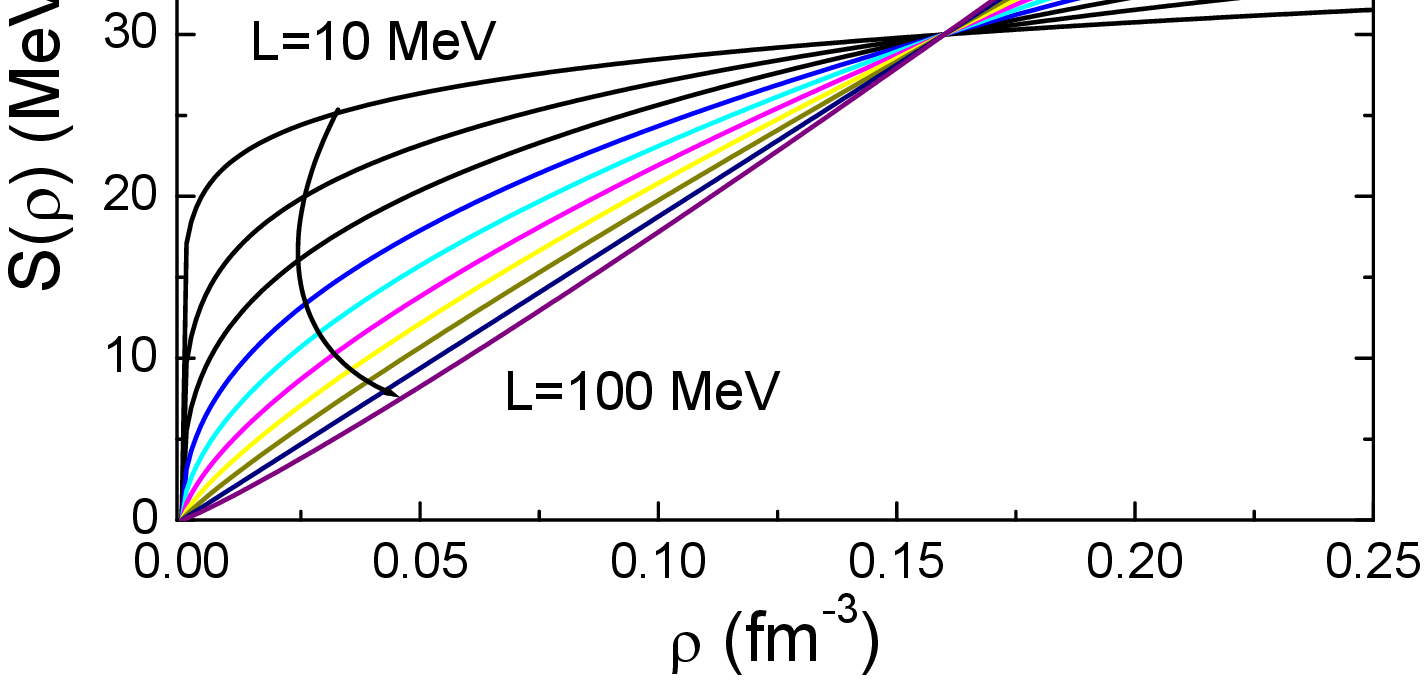}\
\vspace{-3cm} \caption{(Color online) The nuclear symmetry energy $S(\rho)$, defined in Eq.~(\ref{esym-1}), as a
function of the density $\rho$ for various values of the slope
parameter $L$ and  the specific value $J=30$ MeV. } \label{}
\end{figure}


\begin{figure}
\centering
\includegraphics[height=6.1cm,width=8.1cm]{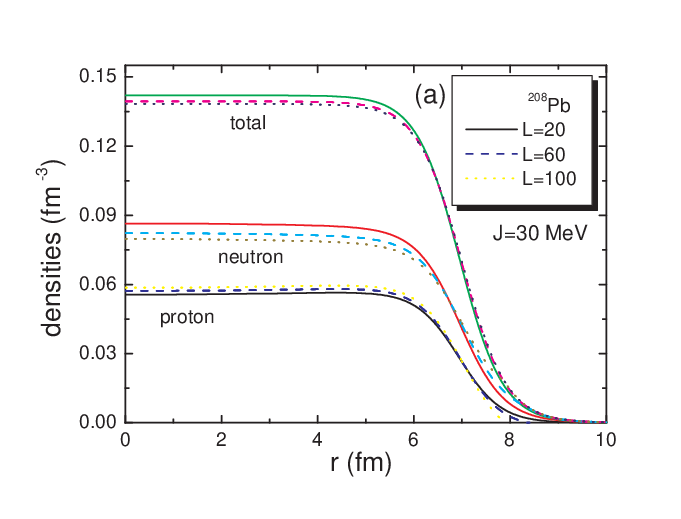}\
\includegraphics[height=6.1cm,width=8.1cm]{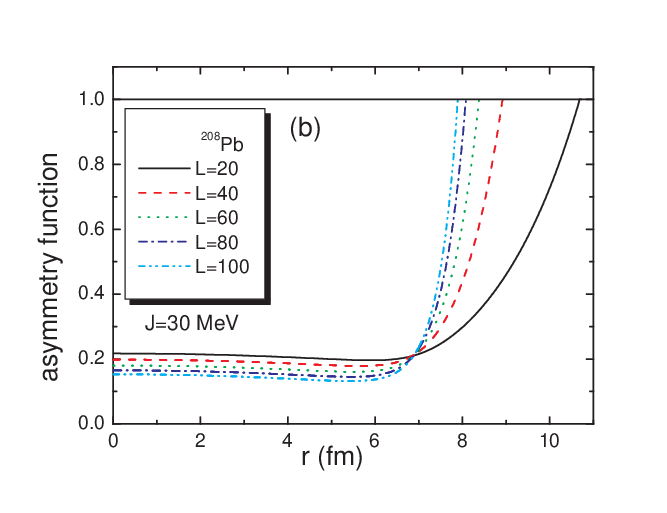}\
\includegraphics[height=6.1cm,width=8.1cm]{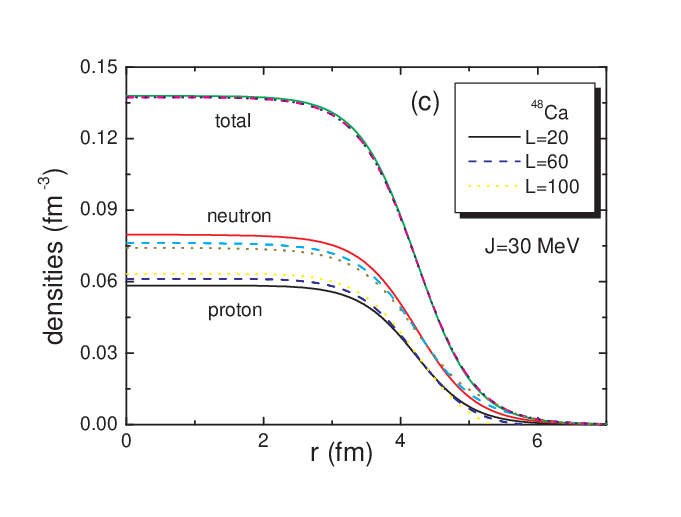}\
\includegraphics[height=6.1cm,width=8.1cm]{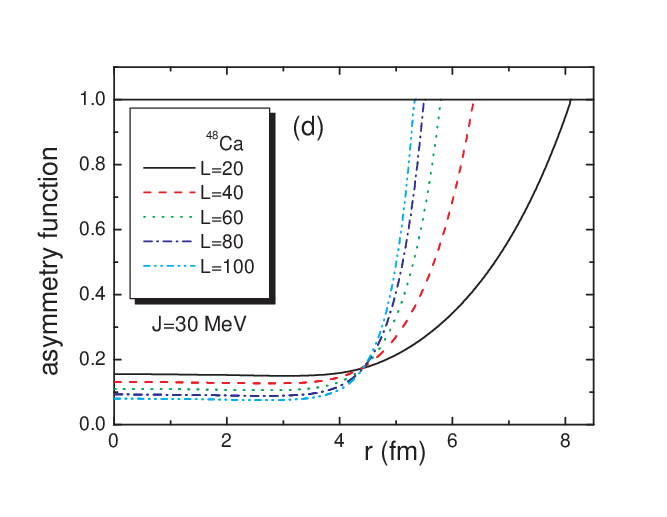}\
\caption{(Color online) The density distribution of neutrons, protons and the
total one (2(a) and 2(c)) for $^{208}$Pb and $^{48}$Ca  for three
values of $L$  (figures 2(a) and 2(c)) and the corresponding
asymmetry functions $\alpha(r)$ for a variety of values of $L$
(figures 2(b) and 2(d)). } \label{}
\end{figure}

\begin{figure}
\centering
\includegraphics[height=6.1cm,width=8.1cm]{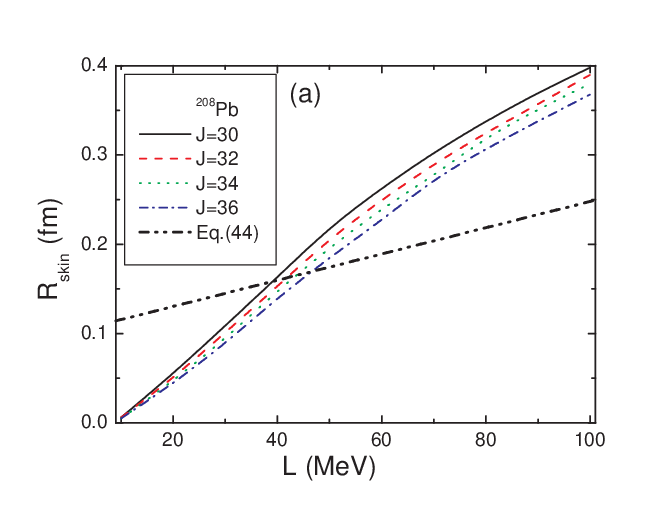}\
\includegraphics[height=6.1cm,width=8.1cm]{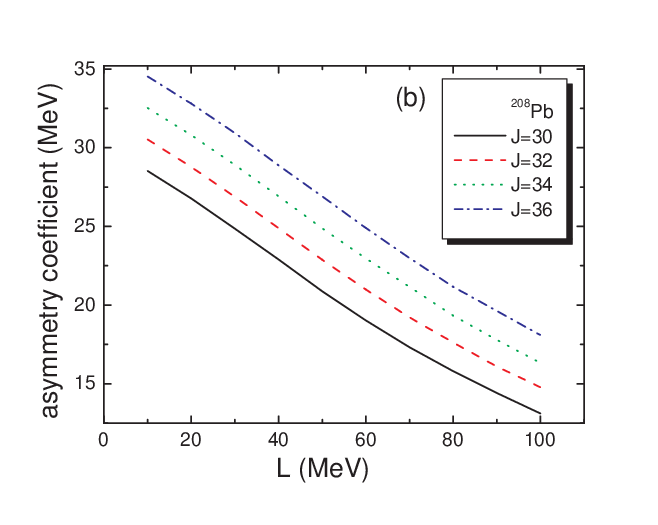}\
\caption{(Color online) (a) The neutron skin $R_{skin}$ for $^{208}$Pb  as a function of the
symmetry energy slope $L$,  for various values of $J$. (b) The asymmetry coefficient  $a_A$ for $^{208}$Pb as a function of the
symmetry energy slope $L$ for various values of the parameter $J$.  } \label{}
\end{figure}

\begin{figure}
\centering
\includegraphics[height=6.1cm,width=8.1cm]{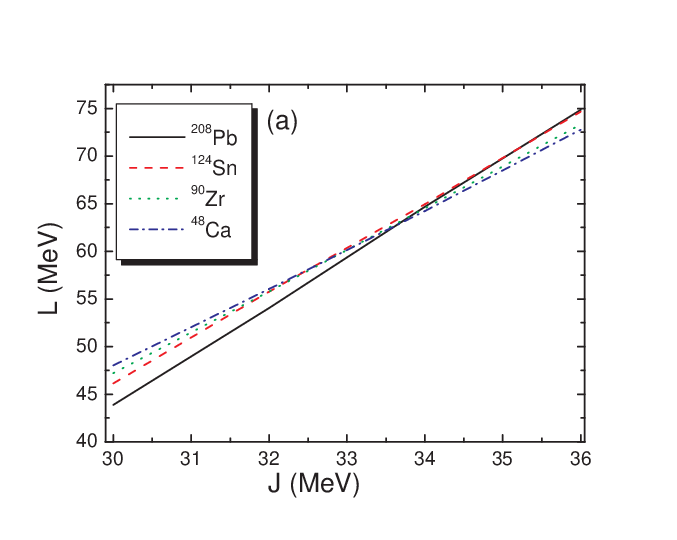}\
\includegraphics[height=6.1cm,width=8.1cm]{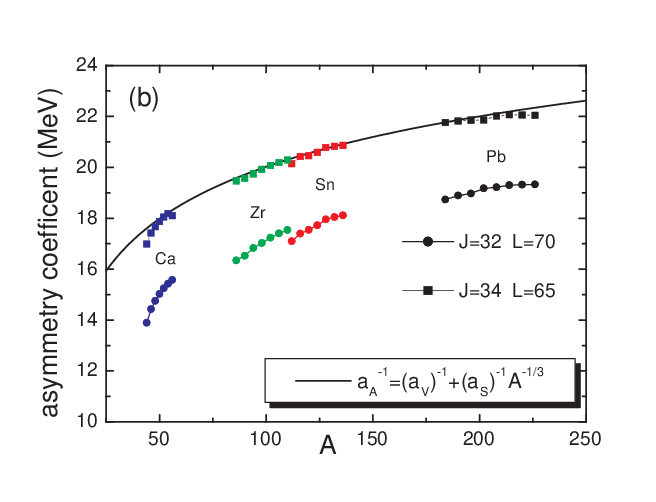}\
\caption{(Color online)(a) The plot of the pairs $L$ and $J$ which reproduce the
empirical value of $a_A$ given by (\ref{fit-Daniel}) for  four
nuclei. (b) The asymmetry coefficients $a_A$  as a function of $A$
for the relevant isotopes and for the set $L=70$, $J=32$ and
$L=65$, $J=34$. The solid line corresponds to the empirical
formula (\ref{fit-Daniel}) (for more details see text).  }
\label{}
\end{figure}

\begin{figure}
\centering
\includegraphics[height=8.1cm,width=9.1cm]{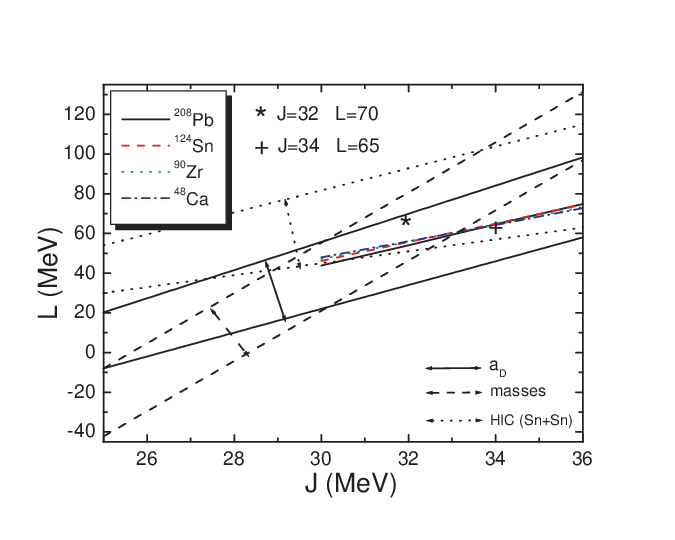}\
\caption{(Color online) Regions of allowed values of pairs $J$ and $L$ (three
bands) constrained from  heavy-ion collisions (HIC(Sn+Sn) case)
and nuclear structure observables ($a_D$ and nuclear masses) (for
more details see Ref.~\cite{Horowitz-14}) in comparison with the
corresponding results constrained from the present approach. The
solid, dashed and dotted arrows indicate constraints related with
$a_D$, nuclear masses  and heavy-ion collisions respectively. The
four colored lines intersecting at the cross show the dependence
of $L$ on $J$ according to Fig.~5(a) of our present work. The star
and the cross correspond to the set $L=70$, $J=32$ and $L=65$,
$J=34$ respectively (for more details see text).
 } \label{}
\end{figure}

\begin{figure}
\centering
\includegraphics[height=6.1cm,width=8.1cm]{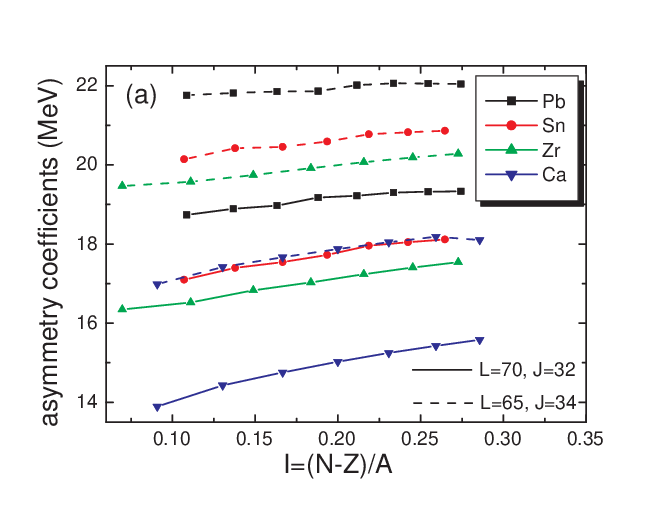}\
\includegraphics[height=6.1cm,width=8.1cm]{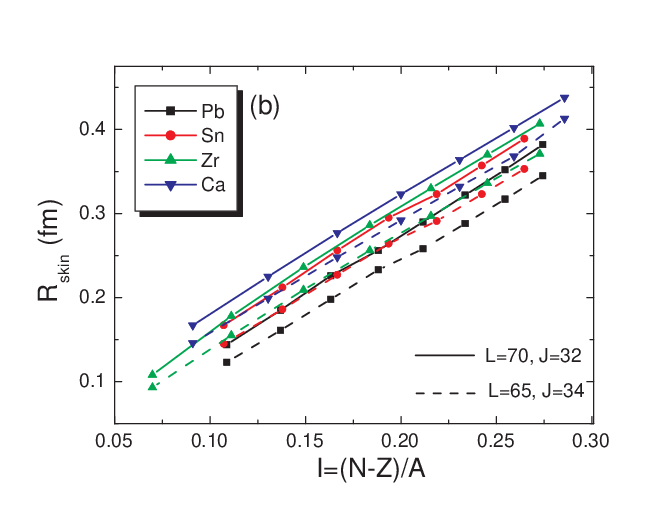}\
\caption{(Color online) (a) The asymmetry coefficient $a_A$ as a function of the
asymmetry parameter $I$ for various isotopes. (b) The
corresponding neutron skin $R_{skin}$ dependence on $I$. }
\label{}
\end{figure}

\end{document}